\documentclass[twocolumn,twocolappendix]{aastex631}

\usepackage{graphicx}%
\usepackage{multirow}%
\usepackage{CJK}

\received{\today}

\submitjournal{AJ}

\shorttitle{Planetary complexity with eigencolours}
\shortauthors{Segal et al.}

\begin{document}

\title{Planetary Complexity Revealed by the Joint Differential Entropy of Eigencolours}

\correspondingauthor{Gary Segal}
\email{g.segal@uq.edu.au, gp.segal@gmail.com}

\author{Gary Segal}
\affiliation{School of Mathematics and Physics,   University of Queensland, St Lucia, Brisbane, QLD, 4072, Australia }
\affiliation{CSIRO Space and Astronomy, PO Box 76, Epping, 1710, NSW, Australia }

\author{David Parkinson}
\affiliation{Center for Theoretical Astrophysics, Korea Astronomy and Space Science Institute Daejeon, 34055, Korea}

\author{Stuart Bartlett}
\affiliation{Division of Geological and Planetary Sciences, California Institute of Technology, Pasadena, CA 91125, USA}



\begin{abstract}

We propose a measure, the joint differential entropy of eigencolours, for determining the spatial complexity of exoplanets using only spatially unresolved light curve data. The measure can be used to search for habitable planets, based on the premise of a potential association between life and exoplanet complexity. We present an analysis using disk-integrated light curves from Earth, developed in previous studies, as a proxy for exoplanet data. We show that this quantity is distinct from previous measures of exoplanet complexity due to its sensitivity to spatial information that is masked by features with large mutual information between wavelengths, such as cloud cover. The measure has a natural upper limit and appears to avoid a strong bias toward specific planetary features. This makes it a candidate for being used as a generalisable measure of exoplanet habitability, since it is agnostic regarding the form that life could take.

\end{abstract}



\section{Introduction} \label{sec:intro}

Thousands of exoplanets have been discovered orbiting other stars \citep{Walker}. Due to their great distances, exoplanets are likely to be observed as spatially unresolved point sources for the foreseeable future \citep{gu2021earth,fan2019earth,jiang2018using}. This presents a challenge of how to measure spatial features and their associated complexity using spatially unresolved time-series data. Being able to infer spatial complexity could be used to identify exoplanets with a potential biosphere while remaining agnostic to the specific composition of that unknown biosphere. 

An agnostic biosignature, such as a generalisable measure of complexity, considers the possibility that alien life may be distinct from life as we know it on earth. Determining the signatures of life is challenging because we have no universal definition of life, we only have the example of life as we find it on earth \citep{barge2022determining,bartlett2020defining,chou2021towards,chou2021planetary,chou_fspas.2021.755100,cleland2019moving,cleland2019quest,guttenberg2021classification,johnson2018fingerprinting,marshall2021}. In addition to seeking molecular or chemical-based signatures (which assumes a fundamentally chemical basis of life), other agnostic approaches focus upon planetary thermodynamic disequilibria \citep{krissansen2016detecting,krissansen2018disequilibrium,krissansen2022understanding,lovelock1965physical,lovelock1975thermodynamics}. Measuring such disequilibria depends critically upon the ability to measure certain compounds in planetary atmospheres as well as careful evaluation of potential abiotic sources.

Researchers have also considered the relationship between life and complexity, broadly defined. The hypothesis that there exists an association between life and the complexity of an exoplanet arises from the fact that living systems exploit information when carrying out complex chemical and physical processes (e.g., \citet{baluvska2016having,davies2016hidden,farnsworth2013living,rashevsky1955life,tkavcik2016information,walker2016informational,witzany2020life}).

Spatially unresolved exoplanet data, in the form of single-point light curves, contains spectral and temporal information generated by simultaneous physical (and possibly biological) processes occurring on the planet. Multiple periodicities, and the mutual information between reflectance variabilities from different wavelength bands, produce complex signals containing potential information about planetary phenomena and their interactions, such as between the biosphere and atmosphere of earth. Signals of this type are not critically dependent on instruments having the spectral resolution or sensitivity to detect specific elements or compounds such as carbon, oxygen or methane. Rather the signal complexity is hypothesised to correlate with the number of distinct processes occurring on a planet or at least the number of distinct spatial features, as would be ascertainable from spatially resolved observation data such as that available for Earth. 

When analysing spatially unresolved exoplanet data, it is entirely possible that complex phenomena occurring on a given planet could be difficult to observe directly. However, metrics such as those described in the present work have the power to extract complex features from within such data. Hence we propose the reflectance signal complexity as an agnostic indicator of planetary complexity, with potential application as a biosignature.

\citet{2022NatAs...6..387B} presented a foundational study using earth observation data from the Deep Space Climate Observatory (DSCOVR) to examine whether time-series reflectance data from multiple wavelength channels, spatially unresolved, 
can be used to measure the complexity of the physical processes occurring on exoplanets. The objective of the study was to address if spatially unresolved time-series data from an exoplanet can be used to infer the presence of a biosphere.

They adopted a metric known as statistical complexity, computed using epsilon machine reconstruction \citep{crutchfield2012between,2022NatAs...6..387B}, and showed that simplified versions of Earth observation data (i.e. earth with clouds and/or certain terrain types removed) correspond to smaller statistical complexity and Shannon entropy values.

\citet{2022NatAs...6..387B} used the average statistical complexity computed across different wavelengths. The results found that this measure discriminated between synthetic earths based first on the presence of cloud cover, where all worlds with clouds returned the largest complexities, and then on multi-surface vs mono-surface worlds, positioning multi-surface worlds without clouds between mono-surface worlds with and without clouds. 

In the present work we extend the analysis of \citet{2022NatAs...6..387B} by introducing an approximation for the entropy of a continuous variable, the Vasicek estimate of the differential entropy \citep{vasicek1976test}. We show that the differential entropy has a natural upper bound of 0 bits for reflectance time-series. We apply principal component analysis (PCA) to the spatially unresolved multi-wavelength data for each synthetic earth used by \citet{2022NatAs...6..387B}, and then calculate the differential entropy for eigencolours deriving the average differential entropy, approximately proportional to the joint differential entropy due to the linear independence of the principal components.

Consistent with past studies \citep{jiang2018using,fan2019earth,gu2021earth,2022NatAs...6..387B} we use DSCOVR measurements to study earth as a proxy exoplanet by using disk-integrated (spatially unresolved) Earth time-series data, with different combinations of spatial features incorporated to construct synthetic worlds at varying levels of spatial complexity. These synthetic worlds, the same as used in \citet{2022NatAs...6..387B}, can provide a reference point for evaluating the spatial complexity of exoplanets using single-point (spatially unresolved) light curves. We present a measure that avoids reliance on any specific wavelength or eigencolour with the objective to avoid over-dependence on any specific spatial feature. Such a measure is more generalisable, providing a potential agnostic biosignature and benchmark for exoplanet habitability.

\section{Data}

We use the same reconstructed Earth data used by \citet{2022NatAs...6..387B}, which consistent with previous studies \citep{jiang2018using,fan2019earth,gu2021earth}, uses disk-integrated point source reflectance data from DSCOVR. The Earth Polychromatic Imaging Camera (EPIC) onboard the DSCOVR provides time-resolved photometric light curves of Earth. The EPIC instrument images the full sunlit disk of Earth using a 2048 × 2048 charge-coupled device (CCD) with 10 narrowband filters centred at 317, 325, 340, 388, 443, 551, 680, 688, 764, and 780 nm. The observations have a temporal resolution of $\approx 68$--$110$ minutes with the data used in this study presented at a temporal resolution of approximately $120$ minutes from 1st January 2016 to 31st December 2016. 

The synthetic Earth data was reconstructed using representative spectra for different spatial features determined by combining the EPIC observations with known ground truth data associated with: cloud cover, vegetation, desert and ocean. The data combined different spatial features to produce reconstructed worlds across a range of reduced complexities by replacing spatial features with a reduced subset, including removing clouds and exposing surface features below. Reconstructed disk images were integrated to single pixels to produce spatially unresolved time-series data, typical of how exoplanets are likely to be observed in the foreseeable future. Further details regarding the production of this data is provided in the original papers \citep{jiang2018using,fan2019earth,gu2021earth,2022NatAs...6..387B}. The reconstructed world data used in this analysis is the same as the data used in \citet{2022NatAs...6..387B}, with the reconstructed worlds and associated world type (groupings) presented in table \ref{tab:tab1}. 

\begin{table}
\centering
\begin{tabular}{ l  l  }
\hline \hline
World type (grouping) & Reconstructed worlds  \\
 & (Earth features retained) \\
\hline 
\multirow{3}{*}{Multi-surface, clouds} & \textbullet{} Original Earth \\
 & \textbullet{} Flora, Ocean, clouds \\ 
  & \textbullet{} Desert, Ocean, clouds \\ 
\hline 
\multirow{3}{*}{Multi-surface, no clouds} & \textbullet{} Original Earth, no clouds  \\
 & \textbullet{} Flora, Ocean, no clouds \\ 
  & \textbullet{} Desert, Ocean, no clouds \\ 
\hline 
\multirow{3}{*}{Mono-surface, clouds} & \textbullet{} Ocean, clouds \\
 & \textbullet{} Flora, clouds \\ 
  & \textbullet{} Desert, clouds \\ 
  \hline 
\multirow{3}{*}{Mono-surface, no clouds} & \textbullet{} Ocean, no clouds \\
 & \textbullet{} Flora, no clouds \\ 
  & \textbullet{} Desert, no clouds \\ 
\hline \hline
\end{tabular}
\caption{Reconstructed worlds used in this analysis}
\label{tab:tab1}
\end{table}

\section{Methods}\label{Method}

\subsection{Differential Entropy}

When only a single image is used to analyse the object of study, be it a galaxy or planet, a distribution over its state space cannot be constructed or even estimated from the data itself. While parts of the image (e.g. distinct measurements or pixels) can in theory be used to derive a finite-sample distribution, such a distribution is not likely to represent the state space of the object of interest, as different parts of the object are distinct. For example plumes, jets and the active nucleus are all distinct parts of the same of object (such as a radio galaxy), and observations of each part cannot be used to infer the distribution of the state space of the galaxy as a whole. This was the case in previous studies where we measured the coarse-grained complexity of images of extragalactic objects using an upper bound approximation of the Kolmogorov Complexity instead of the entropy \citep{2019PASP..131j8007S,10.1093/mnras/stad537}.

On the other hand, if we have a time-series of measurements of the object as a whole, such as a time-series of disk-integrated reflectance data from a planet, we can use this to estimate the state space distribution of disk-integrated reflectance observations given our understanding or assumptions of how the object changes across time. For example, if we take hourly measurements of a planet for five years, collecting only spatially unresolved reflectance data across multiple wavelengths, it seems reasonable that we can infer something about the physical processes occurring on the planet.

In this context, it becomes reasonable to try and estimate the entropy of a continuous variable, such as the disk-integrated reflectance values of an exoplanet, using time-series observations as the sample. We thus propose to estimate the differential entropy using the same data derived by \citet{2022NatAs...6..387B}, except here we treat reflectance, the variable of interest, as continuous with closed support. This makes theoretical sense as depending on the chosen level of precision, the reflectance data can take any value between 0 and 1.

The differential entropy $H(X)$ of a variable $X$ with continuous density function $f(x)$ is defined as \citep{cover1991information}:
\begin{equation}
H(f) = -\int_{\mathcal{X}} f(x) \log f(x) dx,
\end{equation}
where $\int_{-\infty}^{\infty}f(x)=1$ and $f(x)>0$ is the support set $\mathcal{X}$ of $X$.

The intuitive difference between differential entropy and discrete entropy is apparent when considering random samples of a continuous random variable with a uniform distribution and support set $[a,b]=[0,1]$. In this case $f(x) = \frac{1}{b-a} = 1$ and accordingly $H(f) = 0$. 

This is distinct from the discrete Shannon entropy, which will depend on the sample size and bin widths at which one evaluates the probability mass function of discrete observations. With high precision measurements and narrow bins the discrete estimate will be close to the log of the sample size. With wider bins the binning process will treat a larger number of observations as equivalent, losing information, and returning a reduced entropy estimate. Accordingly, the discrete entropy varies depending on the binning and sample size, whereas the differential entropy can be appropriately defined if the continuous density function is known or can be estimated. The differential entropy can naturally provide standardised estimates across different samples with varying sample size and measurement precision (where the probability density function is known). 

Another property of the differential entropy is that it can return negative values unlike the discrete entropy. Consider the general form of the differential entropy for a uniform distribution:
\begin{equation}
H(f) = -\int_{a}^{b} \frac{1}{b-a} \log \frac{1}{b-a} dx = \log(b-a),
\end{equation}
where $a=0$ and $b<1$ the differential entropy will be negative. It is worth noting however that the volume of the support set $2^{\log \frac{1}{b-a}} = \frac{1}{b-a}$ will always be non-negative, as pointed out by \citet{cover1991information}.

As in the discrete case, the differential entropy can also be extended to multiple variables where the conditional entropy can be used to formulate the familiar definition of mutual information \citep{cover1991information}:
\begin{equation}
I(X ; Y) = \int f(x,y)\log\frac{f(x,y)}{f(x)f(y)} dx dy =  H(Y)- H(Y \vert X)\,.  
\end{equation}

The differential entropy can be estimated using the approach introduced by Vasicek in 1976 \citep{vasicek1976test}. Expressing the differential entropy in terms of the derivative of the inverse of the distribution function $F$ from which samples $x_{i}$ are drawn:
\begin{equation}
H(f) = \int_{0}^{1}\log\biggl\{\frac{d}{dp} F^{-1} (p) \biggl\} dp,
\end{equation}
an estimate can be derived by substituting the empirical distribution function for $F$, replacing the differential operator with a difference operator and estimating the derivative of $F^{-1}(p)$ by a function of the order statistics $x_{(i)}$ where $n$ is the sample size and $m$ is the window length for calculating differences:
\begin{equation}
H_{mn} = \frac{1}{n} \sum_{i=1}^{n}\log\biggl\{\frac{n}{2m}(x_{(i+m)}-x_{(i-m)})\biggl\}\,.
\end{equation}

In this paper we use $\log_{2}$ so that the calculated differential entropy will be in bits, consistent with the statistical complexity \citep{2022NatAs...6..387B}.

An open problem concerning Vasicek’s estimator $H_{mn}$ is the selection of a window length $m$ that can take any integer value less than $n/2$ for a given sample size $n$ \citep{vasicek1976test,crzcgorzewski1999entropy,wieczorkowski1999entropy}.  A smaller value of $m$, for fixed $n$, reduces the error resulting from the use of finite differences to approximate the density function. Conversely, errors from approximating the true distribution function $F$ with the empirical distribution function are reduced by smoother approximations returned by a larger value of $m$. Accordingly the optimal choice of $m$, for a given sample size, depends on the (unknown) true distribution function $F$, whereby the smoother the density of $F$ the larger the value of $m$ can be to account for empirical approximation errors. Asymptotically, as $n\to\infty$, the elimination of both sources of error requires that $m\to\infty$ and $m/n\to0$ \citep{vasicek1976test}. In this analysis, consistent with the heuristic formula used elsewhere in the literature, we set the window length to $m = \lfloor{\sqrt{n}+0.5}\rfloor$ \citep{crzcgorzewski1999entropy}.

In the case of the disk-integrated reflectance values of an exoplanet, the support set is bound by $[0,1]$, as it is a measure of the fraction of incident electromagnetic radiation that is reflected. With the support $\mathcal{X}$ being a finite set, the entropy is maximised by a uniform distribution \citep{conrad2004probability}, and accordingly the upper bound of the differential entropy of an exoplanet's reflectance data will be $H(f)=0$. 

\subsection{Joint differential entropy of spectrum eigencolours}

The joint entropy of variables (i.e. reflectance measurements at distinct wavelengths), $H(x,y)$, differs from the sum of entropy values for each wavelength, $H(x) + H(y)$, by the mutual information in $x$ and $y$. 

This yields an insight into how distinct wavelengths or principal components (PCs) from light curves (i.e. the eigencolours of Earth’s spectrum) can be combined. By calculating the sum or average of entropy values for each distinct wavelength, additional weight is placed on the mutual information which contributes more than once. Conversely, the joint entropy $H(x,y)$, or the sum of entropy values for independent components such as when data is transformed as part of principal component analysis (PCA), will count the mutual information only once. 

When determining the entropy of time-series measurements, using the original waveband data, the average entropy is increased by mutual information between wavelengths. In particular the impact of cloud cover on the Earth's reflectance data is apparent across multiple wavelengths \citep{jiang2018using,fan2019earth,gu2021earth}. Take for example wavelengths $\lambda_{1} = 680nm$ and $\lambda_{2} = 780nm$, we use notation that represents the time-series measurements at each wavelength as follows $\mathcal{B}^{\{\lambda_{1}\}} = \mathcal{B}^{\{680nm\}}$. Here each batch of measurements at a specific wavelength constitute a time-series over the same interval $\Delta t$.

The average or sum of time-series values at distinct wavelengths can be expressed with respect to the mutual information $I$:
\begin{equation}
H(\mathcal{B}_{\Delta t}^{\{\lambda_{1}\}})+ H(\mathcal{B}_{\Delta t}^{\{\lambda_{2}\}}) = H(\mathcal{B}_{\Delta t}^{\{\lambda_{1}\}}, \mathcal{B}_{\Delta t}^{\{\lambda_{2}\}}) + I(\mathcal{B}_{\Delta t}^{\{\lambda_{1}\}} ; \mathcal{B}_{\Delta t}^{\{\lambda_{2}\}}) \,.
\end{equation}
The mutual information measured will be large when the entropy for each wavelength is large and there are casual structures between wavelengths (i.e. the impact of cloud cover across multiple wavelengths) that reduce the entropy of $H(\mathcal{B}_{\Delta t}^{\{\lambda_{1}\}})$ by knowing $H(\mathcal{B}_{\Delta t}^{\{\lambda_{2}\}})$:
\begin{equation}
I(\mathcal{B}_{\Delta t}^{\{\lambda_{1}\}} ;\mathcal{B}_{\Delta t}^{\{\lambda_{2}\}}) = H(\mathcal{B}_{\Delta t}^{\{\lambda_{1}\}})- H(\mathcal{B}_{\Delta t}^
{\{\lambda_{1}\}} \vert \mathcal{B}_{\Delta t}^{\{\lambda_{2}\}}) \,.
\end{equation}
The mutual information measured will be small when $H(\mathcal{B}_{\Delta t}^{\{\lambda_{1}\}})$ and $H(\mathcal{B}_{\Delta t}^{\{\lambda_{2}\}})$ are small, since:
\begin{equation} I(\mathcal{B}_{\Delta t}^{\{\lambda_{1}\}} ; \mathcal{B}_{\Delta t}^{\{\lambda_{2}\}}) \le H(\mathcal{B}_{\Delta t}^{\{\lambda_{1}\}})\,. \end{equation}

The mutual information will also be small, however, when there is little causal structure across wavelengths, whereby:
\begin{equation} H(\mathcal{B}_{\Delta t}^{\{\lambda_{1}\}}, \mathcal{B}_{\Delta t}^{\{\lambda_{2}\}}) \approx H(\mathcal{B}_{\Delta t}^{\{\lambda_{1}\}})+ H(\mathcal{B}_{\Delta t}^{\{\lambda_{2}\}})\,.  \end{equation}

In this analysis we then transform the data into time-series of eigencolours using an approach consistent with Gu et al \citep{gu2021earth}. We perform PCA using Singular Value Decomposition (SVD) whereby the reflectance data, in the form of a matrix $\mathbf{R_{[T*\Lambda]}}$ with $T$ rows of mean-centred time step values and $\Lambda$ columns for wavelengths, can be expressed as: 
\begin{equation}
\mathbf{R}_{[T*\Lambda]} = \mathbf{U}_{[T*\Lambda]}\mathbf{\Sigma}_{{[\Lambda*
\Lambda]}}\mathbf{V}^{T}_{[\Lambda*\Lambda]}\,.
\end{equation}

The columns of $\mathbf{V}$ are the set of ordered orthornormal (approximately when implemented)  eigenvectors representing the eigencolours $v$ with rows corresponding to the distinct wavelength columns of $\mathbf{R}$. The columns of $\mathbf{U}$ contain the set of ordered eigenvectors representing the time-series of each eigencolour where each row corresponds to a time step in $\mathbf{R}$. The diagonal matrix $\mathbf{\Sigma}$ contains the singular values of matrix $\mathbf{R}$, representing the distinct square roots of the eigenvalues on which principal components are ordered. 

Irrespective of the mutual information between wavelengths in the original data, the distinct eigencolours $v_{i}$ of the transformed data, corresponding to distinct eigenvalues, will be linearly independent. Accordingly the joint differential entropy of eigencolours (JDE$\Lambda$) will be equivalent to the sum and proportional to the average:
\begin{equation} \label{joint_entropy} H(\mathcal{B}_{\Delta t}^{\{v_{i}\}},..., \mathcal{B}_{\Delta t}^{\{v_{\Lambda}\}}) \propto \frac{1}{\Lambda}\sum_{i}^{\Lambda}H(\mathcal{B}_{\Delta t}^{\{v_{i}\}})\,.  \end{equation}


\section{Results}\label{sec2}

\begin{figure}[h]
    \centering
    \includegraphics[width=0.9\linewidth,height=\textheight,keepaspectratio]{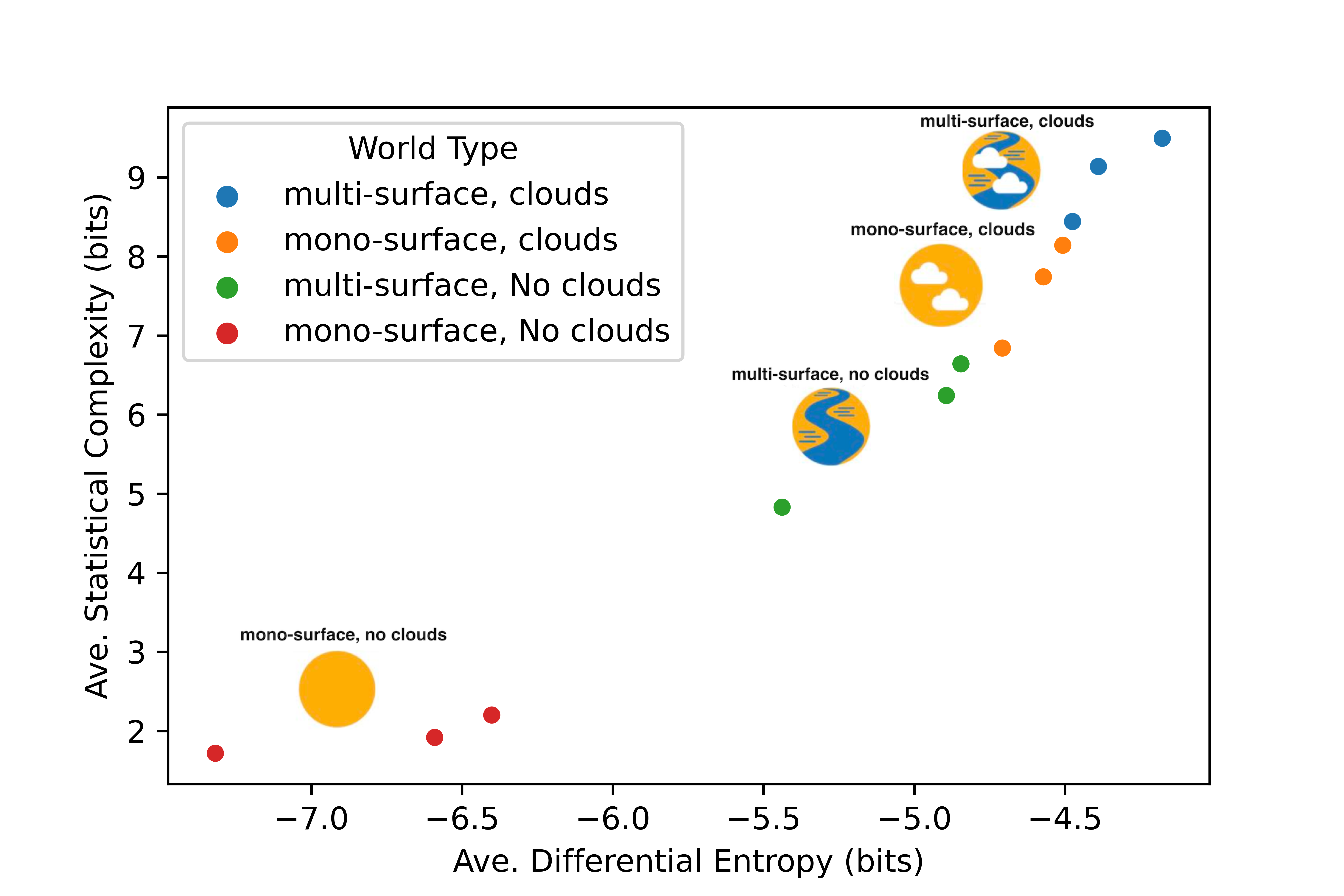}\\
    \includegraphics[width=0.9\linewidth,height=\textheight,keepaspectratio]{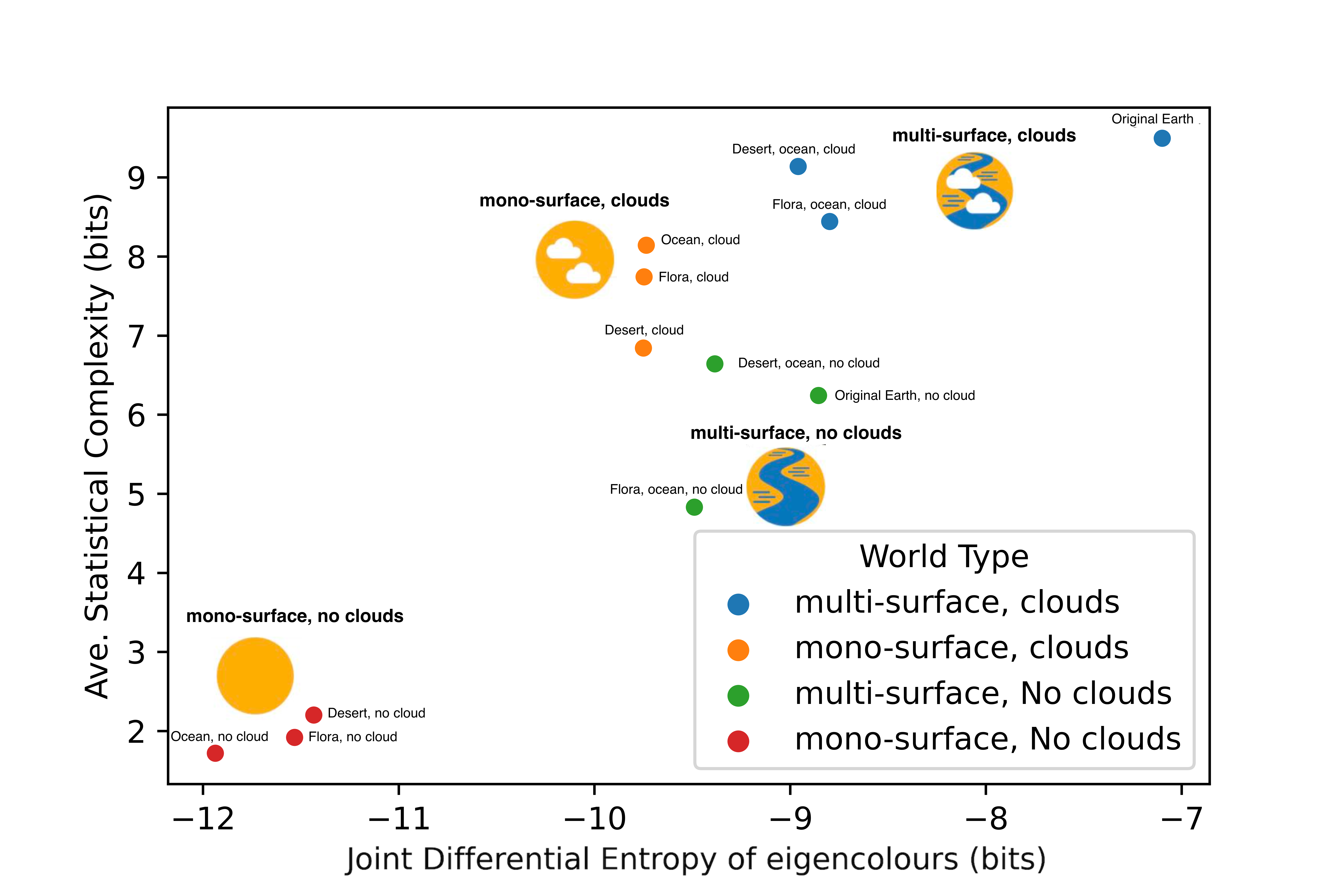}
    \caption{Wavelength average statistical complexity vs. wavelength average differential entropy (\textit{top}) and complexity vs. joint differential entropy of the eigencolours (\textit{bottom}).}
    \label{fig3}
\end{figure}

We computed the principal components and differential entropy values from the same dataset as used by \citet{2022NatAs...6..387B}. 
The wavelength average differential entropy discriminates between the reconstructed synthetic worlds in the same manner as the average statistical complexity computed by \citet{2022NatAs...6..387B}, as shown in \autoref{fig3} (top panel). In this case, consistent with previous findings, cloudy worlds return the largest values. We observe that within the cloudy worlds there is also a relatively small distinction between the mono-surface and multi-surface groups (note the small gap between the most complex mono-surface cloudy world, and the least complex multi-surface cloudy world). It was observed by \citet{2022NatAs...6..387B} that to see any such distinction in the presence of the obscuring effect of clouds was encouraging. 

\begin{table*}
\centering
\begin{tabular}{ l  l  c}
\hline \hline
Description of Complexity & Reconstructed Worlds & \begin{tabular}{l}JDE$\Lambda$ (bits) \end{tabular} \\
\hline 
\begin{tabular}{l} $\geq$ Three surface types \\ and clouds \\  \end{tabular}
& Original Earth & -7.10 \\
\hline 
\multirow{3}{*}{\begin{tabular}{l}$\geq$ Three surface types, \textbf{or} \\ Two surface types and clouds \end{tabular}}
 & Flora, Ocean, clouds & -8.80 \\ 
  & Original Earth, no clouds & -8.85\\
  & Desert, Ocean, clouds & -8.96\\
\hline 
\multirow{5}{*}{\begin{tabular}{l}Two surface types, \textbf{or}\\
one surface type and clouds\end{tabular}}
 & Desert, Ocean, no clouds & -9.38 \\ 
 & Flora, Ocean, no clouds & -9.49 \\ 
 & Ocean, clouds & -9.73 \\
 & Flora, clouds & -9.75 \\ 
 & Desert, clouds & -9.75 \\ 
  \hline 
\multirow{3}{*}{\begin{tabular}{l}One surface type only\end{tabular}} 
  & Desert, no clouds & -11.43 \\ 
  & Flora, no clouds & -11.53 \\ 
  & Ocean, no clouds & -11.94 \\
\hline \hline
\end{tabular}
\caption{Original Earth and reconstructed world data ranked by the Joint Differential Entropy of the eigencolours (JDE$\Lambda$). This ranking can be used as a reference point for future observations of exoplanets.}
\label{tab:benchmark}
\end{table*}

\autoref{fig3} (bottom panel) shows the wavelength average differential entropy replaced with the average differential entropy of the eigencolours derived through PCA (i.e. the principal components of the waveband data used in this and previous studies \citep{gu2021earth, 2022NatAs...6..387B}). 
This approximation of the Joint Differential Entropy of eigencolours is also presented in table \ref{tab:benchmark} where the values are used to rank the Earth and reconstructed worlds. The eigencolours are linearly independent, unlike the original wavelengths, meaning that mutual information across multiple wavelengths, such as that generated by cloud cover, is counted only once. This is demonstrated by the average $R$ (Pearson correlation coefficient) value computed between the 10 eigencolours of approximately (by order of magnitude) $10^{-17}$, for both the original earth data and the reconstructed data. Conversely, when computed between the wavelengths at which the EPIC narrowband filters are centred, the average value returned for the original earth data is $R\approx0.57$ reducing to $R\approx0.08$ for the reconstructed earth data where only cloud cover is removed. This is consistent with observations by \citet{jiang2018using} that reflected light from clouds dominates all 10 EPIC wavelengths. 

\section{Discussion}

The average differential entropy of distinct eigencolours, proportional to the joint entropy, partitions the data primarily on mono-surface and multi-surface groups with cloud cover as a secondary partition. Furthermore, the measure partitions mono-surface and multi-surface worlds even in the presence of clouds, and this distinction is stronger than that made by the wavelength average statistical complexity. This makes the measure complimentary to the wavelength average statistical complexity computed by \citet{2022NatAs...6..387B}, as it avoids the obscuring effect of clouds and reveals a measure aligning with an intuitive notion of complexity, a measure which shows a correspondence with the approximate number of distinct spatial features. As shown in \autoref{fig3} and table \ref{tab:benchmark}, the joint differential entropy of eigencolours (horizontal axis), orders the reconstructed worlds so that worlds with a comparable number of distinct spatial features (i.e. ocean, cloud, desert, vegetation) approximately align. This measure of complexity aligns the original earth with cloud cover removed, with multi-surface worlds where clouds are retained but other surface types are removed (i.e. the flora, ocean, cloud retained but desert and ice replaced with flora). Worlds with only two surface types and no clouds (e.g. desert and ocean, no clouds) remain closer to mono-surface worlds with clouds (e.g. Desert and cloud). In this sense, the measure appears less influenced by any specific spatial feature, potentially making it more generalisable and agnostic on the form that planetary or biological features could take on other planets. Furthermore, unlike the wavelength average statistical complexity (vertical axis), the joint differential entropy of eigencolours (horizontal axis) returns larger values for the original earth data (both with and without clouds) compared to other multi-surface worlds of the same cloud cover type (i.e. with and without clouds), with the original earth data well separated from all reconstructed worlds. This may be due to greater spatial feature diversity and/or nuanced structure present in the original earth data.

\begin{figure}[h] 
\centering
\includegraphics[width=0.9\linewidth,height=\textheight,keepaspectratio]{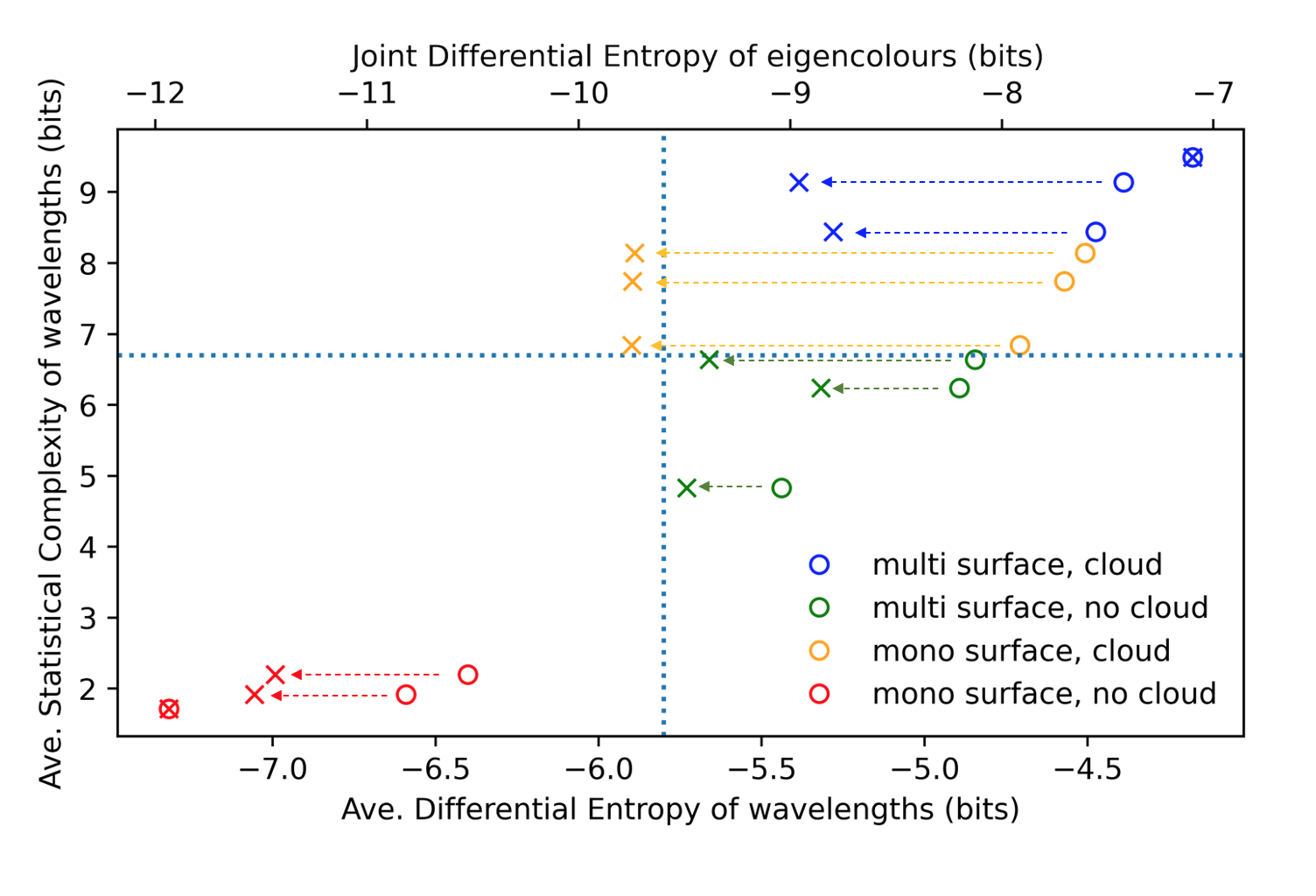}
\caption{\label{fig_dualaxis} A dual horizontal axis is used to plot both the wavelength average differential entropy (circle `$\circ$' shapes), and the joint differential entropy of the eigencolours (`x' shapes), aligning the horizontal axis at the respective values of the original Earth and the least complex synthetic world. The fixed point alignment of the horizontal axes changes the scale of each axis. The vertical axis shows the corresponding wavelength average statistical complexity for both measures. Perforated lines show the distinction between world types.} 
\end{figure}

In \autoref{fig_dualaxis} a dual horizontal axis is used to plot both the wavelength average differential entropy, and the joint differential entropy of the eigencolours, aligning the horizontal axis at the respective values of the original earth and the least complex synthetic world (i.e. with only ocean and no clouds). This chart shows the change in the relative positioning of synthetic worlds with intermediate complexity values as indicated by the arrows. In particular we note the increased relative (re-scaled) distance, in terms of complexity, between mono-surface cloudy worlds and multi-surface cloudy worlds, and the relative increased value of both the original earth and the original earth without clouds when using the joint differential entropy of the eigencolours. 

\autoref{Extended Analysis} provides extended analysis examining both the differential entropy of individual principal components as well as the use of combined eigenvalues (instead of the joint differential entropy). These results show a surprising insight and potential benefit of the joint or average differential entropy of eigencolours for exoplanet characterisation, namely, that where the combined eigenvalues (average or sum) are dominated by cloud cover, the joint differential entropy of eigencolours unmasks a signal for spatial complexity. 
Furthermore, using the joint differential entropy avoids reliance on any specific wavelength or eigencolour. This may avoid a bias toward wavelengths or eigencolours associated with planetary features typical of earth but distinct from those associated with the biospheres of other planets. In this sense the joint differential entropy may provide a more generalisable  measure of complexity compared to the differential entropy associated with a specific eigencolour (such as PC2). The joint measure provides a tool for biosignature detection that is more agnostic on the form a potential exoplanet biosphere may take and provides a potential measure for the future benchmarking of exoplanet complexity as illustrated in table \ref{tab:benchmark}.


\section{Conclusions}

\citet{2022NatAs...6..387B} found that statistical complexity averaged over different wavelengths discriminated between synthetic earths based first on the presence of cloud cover, where all worlds with clouds returned the largest complexities, and then on multi-surface vs mono-surface worlds, positioning multi-surface worlds without clouds between mono-surface worlds with and without clouds. It was a somewhat surprising, but encouraging result that such a distinction between multi-surface and mono-surface worlds 
was possible at all.

In this study we found that the average differential entropy of eigencolours (proportional to the joint differential entropy) provides a powerful measure of spatial feature complexity, despite being computed from spatially unresolved time-series data. The measure appears to 
correlate with the approximate number of distinct spatial features, discriminating between mono-surface and multi-surface worlds even in the presence of clouds. This makes the measure complimentary to the wavelength average statistical complexity values computed by \citet{2022NatAs...6..387B}, as it avoids the obscuring effect of clouds and reveals a measure aligning with an intuitive notion of complexity based on distinct spatial features. A surprising insight is that the 
measure unmasks a signal for surface type complexity 
, despite the fact that principal components associated with cloud cover contribute greater than $83\%$ of the light-curve variability \citep{fan2019earth, gu2021earth}.  

The measure avoids reliance on any specific wavelength or eigencolour and appears to not be overly influenced by any specific spatial feature. This may avoid a bias toward planetary features typical of Earth, making it more generalisable and acting as a potential biosignature approach that is agnostic to the form that life could take on other planets. Furthermore, because the reflectance is bounded between zero and unity, and as such the maximum possible entropy will be generated by a uniform distribution, we can define the upper limit for the differential entropy for any reflectance time-series as zero. The known upper bound of the measure, combined with its generalisability, makes it a potential candidate for being used as an agnostic biosignature and for future applications in determining reference points (benchmarks) of the spatial feature complexity of exoplanets observed as spatially unresolved time-series.

\begin{acknowledgments}
\section{acknowledgments}
GS is supported by CSIRO Space and Astronomy. 
DP is supported by the project \begin{CJK}{UTF8}{mj}우주거대구조를 이용한 암흑우주 연구\end{CJK}, funded by the Ministry of Science. 
SB is supported by NASA Exoplanets Research NNH22ZDA001N-XRP:F.3 2022.
\end{acknowledgments}

%





\bibliography{bibliography}
\bibliographystyle{aasjournal}

\appendix
\section{Extended Analysis}\label{Extended Analysis}

\begin{figure*}[htp]
\centering
\includegraphics[width=0.9\linewidth,height=\textheight,keepaspectratio]{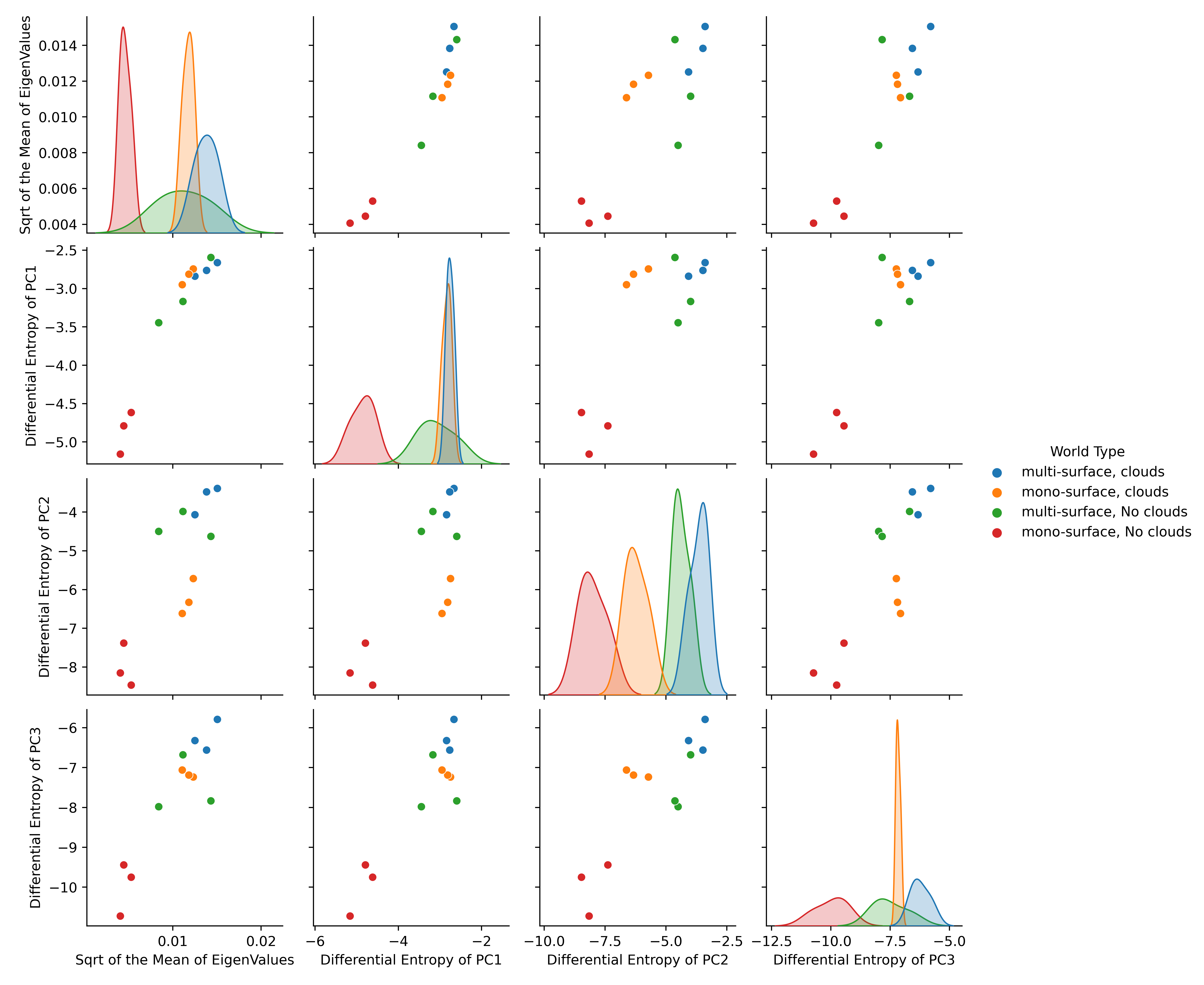}
\caption{\label{fig4} The off-diagonal frames contain pair-wise plots of individual principal components and the wavelength average statistical complexity classified by the world types presented in \autoref{tab:tab1}. The diagonal frames contain overlapping kernel density estimate (KDE) plots for each of the world types.} 
\end{figure*}

Examining the individual principal components it is apparent that PC2 discriminates significantly between mono-surface and multi-surface groups as shown in \autoref{fig4}. This suggests that information about major surface types is found within PC2, in agreement with the findings of \citet{gu2021earth} and \citet{fan2019earth}. We find that PC1 retains a stronger association with cloud cover, also consistent with the regression analysis between eigencolour time-series and spatial feature fractions performed by \citet{gu2021earth}, and the analysis performed by \citet{fan2019earth}. 

\citet{gu2021earth} found that PC1 and PC4 were associated with low and high cloud cover, respectively, and contributed approximately $83\%$ of the Earth's light-curve variability. The differential entropy is distinct from and complimentary to variability measures as analysed by \citet{gu2021earth} and other past studies \citep{fan2019earth,jiang2018using}. \autoref{fig4} shows that the square root of the mean of eigenvalues, or equivalently the mean or sum of eigenvalues, is sensitive to the presence of cloud cover with mono-surface cloudy worlds returning larger values than the original earth without clouds. This is caused by the dominance of the first principal component, that captures information about cloud cover and contributes the majority of light-curve variability (greater than $80\%$ on average across worlds), consistent with the findings of \citet{gu2021earth} and \citet{fan2019earth}. Accordingly the eigenvalue average does not distinguish between mono-surface and multi-surface worlds, in the manner that the joint (average) differential entropy of eigencolours does.
   
\end{document}